\documentstyle[12pt,epsf]{article} 

\newcommand{\address}[1]{\begin{center}#1\\ \end{center}}
\renewcommand{\title}[1]{\begin{center}\Large \bf #1\\ \end{center}}
\renewcommand{\author}[1]{\begin{center}\bf #1\\ \end{center}}
\begin{document}
\begin{flushright}
NORDITA 97/53-P\\
hep-ph/9708386
\end{flushright}
\vspace{1cm}
\title{Ultraviolet dominance of power corrections in QCD?}

\vspace{0.7cm}

\author{V.M.~ Braun}

\address{NORDITA, Blegdamsvej 17, DK--2100 Copenhagen \O, Denmark}

\vspace{1cm}
\address{
Presented at 5th International Conference on Physics Beyond 
the Standard Model, Balholm, Norway, 29 Apr - 4
May 1997} 

\begin{abstract}
I explain  current theoretical ideas on  
higher-twist (hadronization) corrections to physical observables in QCD. 
\end{abstract}

The recent years  witnessed an outburst of theoretical activity devoted to
nonperturbative effects in hard processes in Quantum Chromodynamics. 
For a generic physical observable R one considers the perturbative
expansion complemented by nonperturbative corrections which are expected
to be suppressed by a power of the hard scale Q
\begin{equation}
   R(Q,\underline{x}) = R_0(\underline{x})
         \left[1+ r_1(\underline{x}) \frac{\alpha_s(Q)}{\pi} +
                  r_2(\underline{x}) \left(\frac{\alpha_s(Q)}{\pi}\right)^2 
                +\ldots
             \right] + \frac{A_R^{(p)}(\underline{x})}{Q^{p}}
\end{equation}
Here $\underline{x}$ is a set of dimensionless variables (typically 
momentum fractions), $R_0$ is the tree-level perturbative result, $r_i$ 
give the radiative corrections and $A_R^{(p)}$ is a (nonperturbative)
constant of dimension GeV$^{p}$ 
determining the leading power correction (with
the smallest power of $Q$). Interest in  power suppressed contributions
is due to our desire to make QCD predictions (in the perturbative domain)
as quantitative as possible, which is imperative for a 
precise determination of the parameters of the Standard Model and 
the ongoing searches of ``new physics''.  
The experimental accuracy has reached the level at which
power like effects are observable and can be studied. 
This calls for a broad and systematic study of the theory and 
phenomenology of nonperturbative
effects in hard processes in QCD, for which  little was done
until very recently.

The main obstacle for the phenomenology has been an almost complete 
absence of theoretical predictions, with the Monte Carlo 
hadronization models providing the only approach to the data. 
This situation has changed and we have a ``minimal model'' 
of higher-twist effects which is consistent with QCD, sufficiently 
general, simple enough
and introduces a minimum number of parameters. This model can be directly
confronted with the data; 
both agreement and disagreement would be meaningful
and interesting. This is still not a theory, but an ``educated guess'' 
to shoot at.   
The objective of my talk today is to emphasize  
this recent development which is based on the examination of 
the structure of infrared
contributions in perturbation theory in somewhat  more detail than  
usual in the derivations of celebrated factorization theorems.

To explain the basic idea, I will consider power corrections to
the structure functions of deep inelastic scattering. This example 
provides a benchmark for the theoretical understanding since the 
Operator Product Expansion (OPE) allows to express  
power corrections to moments of the structure functions in terms of 
contributions of higher twist operators. Schematically, one can write
\begin{equation}
  {}\!\!\int\!\! dx\,x^{N-1}F_2(x,Q^2) = C(N,\mu)P(N,\mu) 
    +\frac{1}{Q^2}\sum_k C^{(4)}_k(N,\mu)P^{(4)}_k(N,\mu) 
    +O\!\left(\frac{1}{Q^4}\right)
\label{OPE}
\end{equation}
where $P(N)$ are the moments of quark and gluon parton distributions,
$C(N)$ are the corresponding coefficient functions, while $C^{(4)}_k$ and
$P^{(4)}_k$ are the coefficient functions and matrix elements of
the relevant twist 4 operators, respectively. The summation goes
over the complete set of such operators which are 
rather numerous \cite{twist4}.
   
 {}From the OPE one learns that
the nonperturbative effects are suppressed as $1/Q^2$ 
and are given in terms of universal objects (in fact, moments of 
multiparton correlation functions \cite{MCF}) with  calculable 
coefficients in perturbation theory. Thus, power corrections
to different deep inelastic processes are related; in principle one 
can sacrifice a few measurements to extract  higher-twist distributions 
and then predict the corrections to other processes.
In practice, however,
the multiparton distributions are too complicated;
thus predictive power of the OPE is largerly lost. As a result, despite
being available for 15 years, the results of \cite{twist4,MCF}
on higher twist parton distributions have rarely been used in phenomenological
analysis.

I am  going to formulate a certain approximation and 
start from the observation that the separation of leading twist 
from higher twists is not unique. 
Imagine that the factorization in transverse momenta that is implicit 
in (\ref{OPE}) were implemented by a rigid cutoff $\mu$ such that 
only coeffcients from transverse momenta $k_t>\mu$ were included in 
the coefficient function $C(N,\mu)$.
 The logarithmic cutoff dependence would cancel
between the coefficient function and the parton distribution, as usual.
In addition, one would find, however, a {\it power-like} cutoff dependence
starting from $\mu^2/Q^2$ 
\begin{equation}
  C(N,\mu) = c_0(N) + \frac{\alpha_s}{\pi}\left[c_1(N,\log \mu)+ h_1(N,\log\mu)
 \frac{\mu^2}{Q^2}+\ldots\right]+\ldots
\label{Cpower}
\end{equation}
which is cancelled by higher-twist contributions. This cancellation is 
possible since matrix elements of twist four operators are quadratically 
divergent with the scale $\mu$ serving as an {\it ultraviolet } cutoff and it
means that the coefficient in front of this  quadratic divergence must match
the IR cutoff dependence of the coefficient function 
\begin{equation}
\sum_k C^{(4)}_k(N,\mu)P^{(4)}_k(N,\mu)/ P(N, \log \mu) = 
\mu^2  \frac{\alpha_s}{\pi} h_1(N,\log \mu) + A^{(4)}(N,\log \mu)
\label{Ppower}
\end{equation} 
Thus the leading twist contribution as a whole depends on the prescription 
used to implement the scale separation, just as the separation of the 
leading twist coefficient function and parton distribution does. 
 At first 
sight, such prescription dependence seems to be avoided in the 
$\overline{\mbox{MS}}$ scheme, because power-like dependence on the 
factorization scale $\mu$ does not exist. The problem reappears, however, 
because the  coefficient function now has a factorially divergent 
series expansion in $\alpha_s$ (referred to as infrared renormalon 
divergence). Summing the series again requires a prescription and the 
prescription-dependence is power suppressed precisely as the cutoff 
dependence above. In both cases, the ultraviolet renormalization of 
higher-twist operators must be performed consistently with the definition 
of the leading-twist coefficient function. The sum of leading-twist and 
higher-twist contributions is then unique.

Now comes the central point. As it is well known, the logarithmic scale 
dependence does not in fact cancel exactly in finite-order perturbative 
calculations, and provides a conventional estimate for the accuracy 
of the calculation. This implies a tacit assumption that the scale independent 
constant terms in the next (uncalculated) order of the perturbation theory 
are of the same order of magnitude as scale
dependent logarithmic terms which can easily be reconstructed from the 
requirement that they must exactly cancel the calculated scale dependence
in low orders. Thus, although the scale dependent terms are ultimately 
unphysical and must cancel, their magnitude indicates 
the size of uncalculated higher-order corrections.
Very similarly, the power like scale dependence of perturbation theory
in (\ref{Cpower}) can be used to estimate the size of higher-twist corrections
under the assumption that the ``true'' matrix elements $A^{(4)}$ in    
(\ref{Ppower}) are roughly proportional to their quadratically divergent
pieces
\begin{equation}
 A^{(4)}(N,\log \mu) \simeq  \mu_0^2   h_1(N,\log \mu)
\label{UVD} 
\end{equation}
where $\mu_0$ is a certain scale of order $\Lambda_{\rm QCD}$.
Note that $\mu_0$ is assumed to be (approximately) independent of N;
Thus, the N dependence of higher-twist corrections is tied up with
(calculable) N dependence of IR contributions in perturbation theory.
I will refer to this assumption as ultraviolet dominance of power 
corrections \cite{BBM97}.

Since  the coefficient function $C$ in (\ref{OPE}) does not depend 
on the target (by definition), so does the  ratio of the higher twist
correction to the leading twist parton distribution in (\ref{Ppower}).
The ultraviolet dominance approximation should be viewed, therefore,  as
a ``minimal model'' 
of supposedly large generic target-independent higher twist corrections, 
providing an ``irreducible background'' for ``true'' effects of quark-gluon
nonperturbative correlations. Note that from the
phenomenological point of view the higher-twist correction can only be 
defined as a sum of ``true'' nonperturbative effects and perturbative 
corrections beyond available order. 
 
Calculation of the characteristic functions  $h_1(N)$ in leading-order 
radiative corrections is relatively easy and can be done using 
different techniques \cite{beneke93,BBZ94,BBB95,DMW96}.
The results exist  for all flavor non-singlet deep inelastic 
structure functions 
and  for the polarized structure function $g_1$ 
\cite{DMW96,HMSSM96,DW96}. A preliminary comparison to the existing data  
was done in \cite{MSSM97} (see also \cite{DMW96}), with encouraging results.
Similar calculations for flavor-singlet contributions are in progress
\cite{mankiewicz}.

Extension of these results beyond leading order faces difficulties,
both techical and conceptual, which I do not have time to discuss in this
talk.

Since the approach is purely perturbative it can be generalized to 
a variety of other processes where the OPE does not exist. 
The inclusive cross section of hadron
production in $e^+e^-$ annihilation \cite{DW97,BBM97} provides 
an instructive example. Tracing IR contributions $\sim \mu^2/Q^2$ in the 
leading twist coefficient function we get an ansatz for the power 
corrections to the longitudinal and total cross sections \cite{BBM97} 
\begin{eqnarray}
\frac{d\sigma_L^{\rm power}}{dx}(x,Q^2) &=& 
\frac{\mu_0^2}{Q^2}\int_x^1\frac{dz}{z}\,
\Bigg\{c_{q,L}\left[\delta(1-z)+\frac{2}{z}
\right]\,D_q(x/z,\mu) \nonumber\\
&&\,+ c_{g,L}\,\frac{1-z}{z^3}\,D_g(x/z,\mu)
\Bigg\}~,
\label{parL}
\\
\frac{d\sigma_{L+T}^{\rm power}}{dx}(x,Q^2) &=& 
\frac{\mu_0^2}{Q^2}\int_x^1\frac{dz}{z}\,
\Bigg\{c_{q,L+T}\left[-\frac{2}{[1-z]_+} + 1 + 
\frac{1}{2}\delta'(1-z)\right]\,D_q(x/z,\mu) 
\nonumber\\
&&\,+ c_{g,L+T}\,\frac{1-z}{z^3}\,D_g(x/z,\mu)
\Bigg\}~,
\label{parLT}
\end{eqnarray} 
where $x$ is the energy fraction carried by the detected hadron and 
$D_{i}$ denotes the leading-twist fragmentation function for parton 
$i=q, g$. The parametrization depends on four constants $c_q$, $c_g$ 
which have to be fitted from the data or related to some nonperturbative
parameters (see below). With the overall scale factor $\mu_0 =1$~GeV these
constants are expected to be of order unity. One can check \cite{BBM98}
that the expressions in (\ref{parL}), (\ref{parLT}) correspond to 
ultraviolet contributions to multiparton fragmentation functions introduced
in \cite{BB91}.   

The most interesting potential application of these methods is to
hadronic event shape observables. The major result in this case 
is that nonperturbative corrections to generic observables are large,
of the order $1/Q$ \cite{W94,DokW95,KS95,AZ95}, and are due to
soft gluon emission.  Since the soft emission exponentiates, so does
the leading nonperturbative correction \cite{KS95}. This leads to 
specific predictions for event shape distributions \cite{DokW97} which
appear to be in agreement with the data.

Most of the specific predictions have been obtained so far using the 
concept of the universal effective coupling 
\cite{DokW95,DMW96}. In this language nonperturbative 
corrections to physical cross sections are due to nonperturbative 
contributions to the effective coupling  and are parametrized
by its moments. In practice, application of this technique amounts to
choosing the same IR parameter $\mu_0^2$ in the relation (\ref{UVD})
for all physical observables (but different for  
corrections of different power). This implies universality of nonperturbative 
scales for Euclidian and Minkowskian observables which would be very 
interesting to test experimentally.

To summarize, there are ongoing efforts to understand the structure of
nonperturbative corrections to hard processes in QCD by using their 
relation to IR effects  in perturbation theory.
In case that the operator structure of the higher twists is known,
these corrections can be identified as being due to ultraviolet regions
in the relevant matrix elements, which explains the title of my talk.
The resulting expressions should be viewed as theoretically
motivated parametrizations of higher twist corrections by a few 
phenomenological parameters. 

A striking feature of these parametrizations is that they are 
target-independent, if weighted by leading twist distributions as 
explained above. This approximation tacitly assumes absense of 
any enhancements of higher twists due to strong quark-gluon
correlations in hadrons and is supported by estimates 
of higher twist corrections to deep inelastic sum rules 
(see a discussion in \cite{braun95}) which appear to be of the same 
order of magnitude as renormalon ambiguities in perturbation theory. 
It is also supported by the recent analysis in \cite{KKPS97}
which suggests that phenomenological extractions of higher twist corrections 
essentially parametrize higher order perturbative rather than nonperturbative
effects. 
Consider, however, deep-inelastic scattering at small x 
from a heavy nucleus.   The 
main higher twist corrections are
due in this case to Glauber-type rescattering effects which have nothing
to do with the leading twist perturbation theory. The ultraviolet
dominance approximation 
would miss the rescattering effects (and thus the A dependence) completely.
The reason is precisely that the large number of nucleons 
provides a new parameter which governs the hierarchy of higher twist effects.

The moral is that the significance  of target-dependent vs.
target-independent higher twist effects has to be studied experimentally,
which provides a very particular task for the phenomenology.
On the other hand, the theory in its present form is still far from 
being complete. One problem is that IR contributions to gluon fragmentation 
are not calculated directly, but are related to quark contributions
of particular type (chains of bubbles) or to the dependence on a certain
naive IR regulator like the gluon mass. This identification is not exact 
for insufficiently inclusive observables, which causes ambiguities 
in predictions, see \cite{BBM97} for a detailed discussion.
This is related to the observation that universality of IR contributions 
for event shape 
observables is spoiled by emission of soft gluons at large angles which
is process-dependent \cite{NS95}. 
A more general problem is that the strong coupling in (\ref{Ppower}) stands
at a low scale $\mu_0$ which indicates that all orders of perturbation
theory are formally important (unless one can prove the opposite by
calculation of the anomalous dimension of the relevant higher twist 
operators). A discussion of these (and other) difficulties as well as 
further theoretical ideas on  power corrections and related issues 
can be found in recent reviews \cite{reviews}.

\end{document}